\documentclass[aps,prl,reprint,twocolumn,superscriptaddress,floatfix,nofootinbib,longbibliography]{revtex4-1}
\usepackage{graphicx,amsmath,amsfonts,amssymb,amsthm,xr}
\usepackage{epsfig,amsmath,amssymb,color,dsfont,upgreek,physics}
\usepackage{mathrsfs}
\usepackage{mathtools}
\usepackage{bbold}
\usepackage{comment}
\usepackage{float}
\usepackage[bookmarks=true,colorlinks,linkcolor=OrangeRed,urlcolor=NavyBlue,citecolor=RoyalBlue]{hyperref}
\usepackage[dvipsnames]{xcolor}
\usepackage{orcidlink}

\definecolor{mygold}{rgb}{0.93,0.69,0.13}

\definecolor{mypurple}{rgb}{0.49,0.18,0.56}

\renewcommand{\H}{\hat{\mathcal{H}}}
\newcommand{\hc}{\text{H.c.}}
\newcommand{\Zt}{$\mathbb{Z}_2$ }

\newcommand{\ad}{\hat{a}^{\dagger}}
\renewcommand{\a}{\hat{a}}
\newcommand{\tauZ}{\hat{\tau}^{z}}
\newcommand{\tauX}{\hat{\tau}^{x}}

\renewcommand{\le}{\left(}
\renewcommand{\r}{\right)}

\begin{document}
\title{Confinement in 1+1D \texorpdfstring{$\mathbb{Z}_2$}{Z2} Lattice Gauge Theories at Finite Temperature}

\author{Matja\v{z} Kebri\v{c}${}{\orcidlink{0000-0003-2524-2834}}$}
\email{matjaz.kebric@physik.uni-muenchen.de}
\affiliation{Department of Physics and Arnold Sommerfeld Center for Theoretical Physics (ASC), Ludwig-Maximilians-Universit\"at M\"unchen, Theresienstra\ss e 37, D-80333 M\"unchen, Germany}
\affiliation{Munich Center for Quantum Science and Technology (MCQST), Schellingstra\ss e 4, D-80799 M\"unchen, Germany}

\author{Jad C.~Halimeh${}{\orcidlink{0000-0002-0659-7990}}$}
\email{jad.halimeh@physik.lmu.de}
\affiliation{Department of Physics and Arnold Sommerfeld Center for Theoretical Physics (ASC), Ludwig-Maximilians-Universit\"at M\"unchen, Theresienstra\ss e 37, D-80333 M\"unchen, Germany}
\affiliation{Munich Center for Quantum Science and Technology (MCQST), Schellingstra\ss e 4, D-80799 M\"unchen, Germany}

\author{Ulrich Schollw\"ock${}{\orcidlink{0000-0002-2538-1802}}$}
\affiliation{Department of Physics and Arnold Sommerfeld Center for Theoretical Physics (ASC), Ludwig-Maximilians-Universit\"at M\"unchen, Theresienstra\ss e 37, D-80333 M\"unchen, Germany}
\affiliation{Munich Center for Quantum Science and Technology (MCQST), Schellingstra\ss e 4, D-80799 M\"unchen, Germany}

\author{Fabian Grusdt${}{\orcidlink{0000-0003-3531-8089}}$}
\email{fabian.grusdt@physik.uni-muenchen.de}
\affiliation{Department of Physics and Arnold Sommerfeld Center for Theoretical Physics (ASC), Ludwig-Maximilians-Universit\"at M\"unchen, Theresienstra\ss e 37, D-80333 M\"unchen, Germany}
\affiliation{Munich Center for Quantum Science and Technology (MCQST), Schellingstra\ss e 4, D-80799 M\"unchen, Germany}

\begin{abstract}
Confinement is a paradigmatic phenomenon of gauge theories, and its understanding lies at the forefront of high-energy physics.
Here, we study confinement in a simple one-dimensional \Zt lattice gauge theory at finite temperature and filling, which is within the reach of current cold-atom and superconducting-qubit platforms.
By employing matrix product states (MPS) calculations, we investigate the decay of the finite-temperature Green's function and uncover a smooth crossover between the confined and deconfined regimes.
Furthermore, using the Friedel oscillations and string length distributions obtained from snapshots sampled from MPS, both of which are experimentally readily available, we verify that confined mesons remain well-defined at arbitrary finite temperature.
This phenomenology is further supported by probing quench dynamics of mesons with exact diagonalization.
Our results shed new light on confinement at finite temperature from an experimentally relevant standpoint.
\end{abstract}

\date{\today}
\maketitle

\textbf{\textit{Introduction.---}}
Lattice gauge theories (LGTs) were first proposed to unravel the intricate mechanism of quark confinement \cite{Wilson1974}, which is one of the key steps towards understanding the formation of hadrons at finite temperature and their transition to quark-gluon plasma \cite{Kapusta2006}.
Although LGTs are still mainly considered when tackling problems in high energy physics, they are also extremely powerful when applied to condensed matter physics \cite{Wegner1971, Kogut1979, Wen2004}.
There, confined phases emerge in many models which are used to describe strongly correlated systems \cite{Sedgewick2002, Sachdev2016}, and \Zt LGTs have direct connections to high-$T_c$ superconductivity \cite{Lee2007, Senthil2000}.
LGTs' full power is unveiled when gauge fields are coupled to dynamical matter at finite doping, where the confinement--deconfinement transition still lacks a comprehensive theoretical description.
This is also partially due to the fact that numerical simulations of LGTs are demanding \cite{Banuls2017}, especially when the dimension surpasses the simplest case of one spatial and time dimension ($1+1$D) \cite{Magnifico2021}.
The study of LGTs becomes even more involved at finite temperature, where the usual numerical limitations are amplified. 

Significant advances in quantum simulations using cold atoms in recent years introduced a new platform to study strongly correlated many-body problems \cite{Greiner2002, Bloch2008, Bloch2012}.
Considerable progress has been made specifically towards quantum simulation of LGTs using cold atoms \cite{Aidelsburger_2021}.
A first proof of concept of experimentally simulating a \Zt LGT has already been made \cite{Schweizer2019, Goerg2019} by employing a Floquet scheme \cite{Barbiero2019}.
Recently, new proposals have been put forward that utilize Rydberg tweezer arrays \cite{Homeier2023}, where the tedious implementation of the gauge protection has been greatly simplified by making use of the so-called local pseudogenerators \cite{Halimeh2022DisorderFree, Halimeh2022LPG}.
Furthermore, proposals using superconducting qubits have also appeared \cite{HomeierPRB2021}.
A lot of effort has also been made in using digital quantum computers \cite{Zohar2017, Irmejs2022, PardoGreenberg2023, Davoudi2022_LGT, Fromm2023} with a version of a LGT already experimentally realized \cite{Mildenberger2022}, however, limited in size.

Here we study finite-temperature properties of a simple $1+1$D \Zt LGT where dynamical charges are coupled to a gauge field at finite doping.
This \Zt LGT is the simplest non-trivial, LGT which can be obtained after discretization of the $U(1)$ Schwinger model to a $\mathbb{Z}_n$ LGT \cite{Ercolessi2018}, and is already within the reach of existing quantum simulators \cite{Aidelsburger_2021, Schweizer2019, Goerg2019, Barbiero2019}.
The dynamics of the gauge field is induced by an electric-field term, which also acts as a linear confining potential in the sector without background charges.
As a result, individual particles become confined into mesons which themselves remain dynamical.
So far, the study of confinement in a \Zt LGT at finite temperature has been limited to challenging Monte Carlo calculations \cite{Grusdt2020}, and the sign problem could be mitigated in a $U(1)$ Schwinger model \cite{Gattringer2015, Gattringer2015U1Higgs, Gattringer2016}.
A theoretical study of a phase diagram at finite temperature and chemical potential utilizing digital quantum simulator algorithms has also been performed, however the study of confinement was hindered by small system size \cite{Davoudi2022_LGT}.

In this work, we employ large scale state-of-the-art matrix product states (MPS) calculations \cite{Schollwoeck2011}, where we make use of the concept of quantum purification \cite{Feiguin2005, Zwolak2004, Nocera2016} in order to obtain finite-temperature states.
We study the decay of the $\mathbb{Z}_2$-invariant Green's function at finite temperature, which is a direct probe of confinement, and uncover a smooth confinement--deconfinement crossover at finite temperature.
This goes against the conventional wisdom where one would expect a deconfined phase at any finite temperature $T > 0$, since the system has to be deconfined in the limit $T \rightarrow \infty$.
In addition to the Green's function, we study Friedel oscillations, which also contain direct signatures of confinement.
Furthermore, we sample snapshots from MPS and study string and anti-string length histograms, that we propose as a new simple but robust measure of confinement suitable for cold-atom experiments.
\begin{figure*}[t]
\epsfig{file=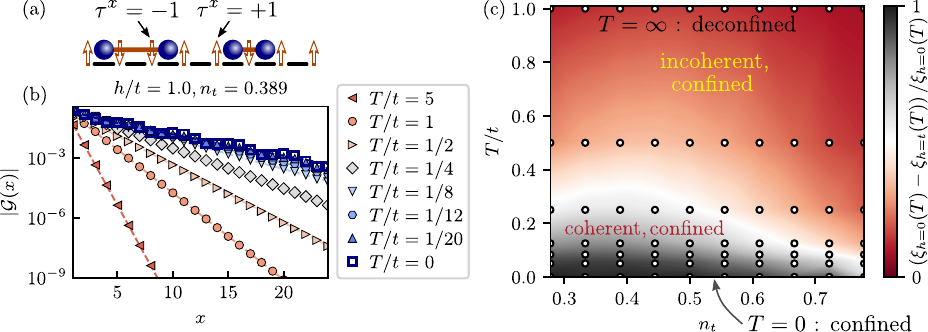}
\caption{(a) In the physical sector without background charges, pairs of hard-core bosons (blue spheres) are connected with the \Zt strings (red horizontal lines), which denote the orientation of the \Zt electric field.
(b) Green's function \eqref{eqGreensFct} for different temperatures $T$ at a constant chemical potential yielding a filling of $n_t = 0.389$ in the ground state,  together with the fitting function containing exponential and power-law decay (dashed lines) \cite{SMPhysRev}.
(c) Heat diagram of the difference between the correlation lengths at $h = 0$ and $h = t$ as a function of target lattice filling $n_t$ and temperature $T$ (white dots indicate data points).
For the details on exact fillings at finite temperatures see \cite{SMPhysRev}.
}
\label{figOne}
\end{figure*}
All these quantities, as well as dynamical quenches at finite and zero temperature, show signatures of confinement at any temperature $T < \infty$, albeit becoming less pronounced as $T$ increases.

\textbf{\textit{Model.---}}
We consider a $1+1$D \Zt LGT where hard-core bosons (partons) are minimally coupled to a \Zt gauge field
\cite{Prosko2017, Borla2020PRL, Kebric2021, KebricNJP2023}
\begin{equation}
    \H = -t \sum_{j} \le \ad_{j} \tauZ_{j, j+1} \a_{j+1} +\hc \r
    - h \sum_{j} \tauX_{j, j+1}.
    \label{eqLGTHamiltonian}
\end{equation}
Here $\ad_{j}$ $(\a_{j})$ are hard-core boson creation (annihilation) operators, and we represent the \Zt gauge and electric fields on the links between lattice sites with Pauli matrices $\tauZ_{j, j+1}$ and $\tauX_{j, j+1}$.
We note that in 1+1D one can map the bosons to fermions via the Jordan-Wigner transformation \cite{Jordan1928}, meaning that our results can be also extended to spinless fermions.

\begin{figure*}[ht]
    \centering
    \epsfig{file=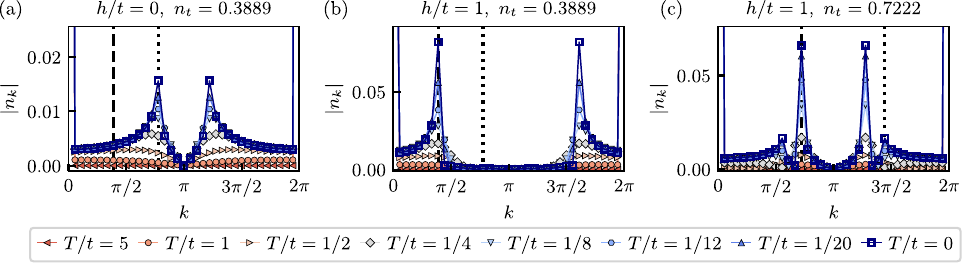}
    \caption{Fourier transformation of Friedel oscillations.
    (a)~Fourier coefficients $|n_k|$ in the deconfined phase exhibits broad peaks at $k = 2 \pi n_t$ (vertical dotted line), which  correspond to Friedel oscillations of free partons.
    (b)~In the confined phase the Fourier transformation exhibits peaks at $k = \pi n_t$ (vertical dashed line), which correspond to Friedel oscillations of mesons.
    (c)~Substantial peaks are visible at $k = \pi n_t$ (vertical dashed line) and at $k = 2 \pi n_t$ (vertical dotted line) at higher target filling $n_t = 26/36$ in the confined phase $h/t = 1$.
    Both peaks rise simultaneously with decreasing temperature $T$. For precise fillings at finite temperature see \cite{SMPhysRev}.
    }
    \label{figTwo}
\end{figure*}

In addition, we consider the set of local operators \cite{Prosko2017}
\begin{equation}
    \hat{G}_j = \tauX_{j-1, j} \tauX_{j, j+1} (-1)^{\hat{n}_j},
    \label{eqGausssLaw}
\end{equation}
where $\hat{n}_j = \ad_j \a_j$.
These local operators generate the local symmetry of the \Zt gauge group and are the \Zt LGT counterpart of the Gauss law.
They commute with the Hamiltonian, $\big [ \H, \hat{G}_{j} \big ] = 0, \forall j$, and with each other, $\big [ \hat{G}_{j}, \hat{G}_{i} \big ] = 0$.
The eigenvalues of $\hat{G}_j$ are $ g_j = \pm 1$.
The Hilbert space can thus be divided into different sectors specified by the values of $g_j$ on each lattice site.
In this work we choose the so-called physical sector without background charges where $g_j = 1, \forall j $ \cite{Prosko2017}.
Hence, the orientation of the \Zt electric field changes only across an occupied lattice site and it is thus convenient to define the \Zt electric \textit{string} and \textit{anti-string}, which graphically represent the orientation of the electric field as $\tau^x =\mp1$, respectively; see Fig.~\ref{figOne}(a).
We note that we do not include a staggered mass term in our LGT, which would give the vacuum state as the ground state in the Schwinger model \cite{Wilson1974, Ercolessi2018}.
This is because we are interested in finite fillings, which translates to finite hole doping in a $t-J_z$ model to which the above LGT can be exactly mapped \cite{Kebric2021}.

The first term in Hamiltonian \eqref{eqLGTHamiltonian} is the hopping term where the $\tauZ$ operator ensures that the Gauss law remains satisfied, i.e., that the partons remain attached to a string.
The second term induces a linear confining potential among partons connected with the same string, since strings become energetically unfavorable.
In the ground state, partons connected with the same string thus become confined into mesons (dimers), where the string length is minimized.
This happens for any non-zero value of $h>0$ \cite{Borla2020PRL}; 
at $h = 0$ partons are free/deconfined \cite{Prosko2017}.
A solution of the confinement problem in the ground state of this \Zt LGT has been found by performing a non-local transformation to the so-called string-length basis \cite{Kebric2021}.
There, confinement can formally be understood as translational-symmetry breaking in the new basis \cite{Kebric2021}.

We use the concept of quantum purification \cite{Feiguin2005, Zwolak2004, Nocera2016, Feiguin2013} in order to obtain finite-temperature states.
We add an auxiliary lattice site to every physical lattice site.
These are entangled to the physical lattice sites and act as a thermal bath \cite{Feiguin2005}.
By using DMRG \cite{Schollwoeck2011, White1992}, we first compute the maximally entangled state between the auxiliary and physical sites on which we then perform imaginary time evolution \cite{Paeckel2019} in order to obtain states at finite temperature $T$ \cite{Feiguin2005, Zwolak2004, Nocera2016}.
We use \textsc{SyTen} \cite{hubig:_syten_toolk, hubig17:_symmet_protec_tensor_network}, an MPS toolkit where DMRG as well as standard time evolution algorithms for MPS are implemented.
For more details on the numerical calculations see \cite{SMPhysRev}.

For practical purposes, we consider an even number of hard-core bosons in the lattice.
Since we employ open boundary conditions, we consider that the chain always starts with an anti-string, i.e., a link with positive orientation $\tau^{x}_{0, 1} = +1$ in the confined phase when $h/t > 0$.
These conditions prevent the partons from being confined to the boundaries.
This is automatically satisfied in the numerical implementation with DMRG in the ground state, where we map the model to a spin-$1/2$ system and also add a chemical potential term proportional to $\mu$ \cite{SMPhysRev}.

\textbf{\textit{Green's function.---}}
In order to probe the confinement of partons into mesons we consider the $\mathbb{Z}_2$-invariant Green's function defined as \cite{Borla2020PRL, Kebric2021, KebricNJP2023}
\begin{equation}
    \mathcal{G}(i-j) = \left \langle \ad_i 
    \le \prod_{i \leq \ell < j} \tauZ_{ \ell, \ell + 1} \r \a_j \right \rangle ,
    \label{eqGreensFct}
\end{equation}
which can also be considered as a one-dimensional version of the Fredenhagen-Marcu order parameter \cite{Gregor2011}.
At ${T = 0}$, it decays exponentially in the confined regime and with a power-law in the deconfined regime \cite{Borla2020PRL}.

The Green's function decays exponentially in both regimes at $T>0$, albeit with different decay rates.
This makes a clear distinction between the confined and deconfined phases at finite temperature difficult.
To overcome this complication, we compare the rate of decay of the Green's function \eqref{eqGreensFct} in both regimes and determine the crossover temperature, at which the thermal fluctuations start to dominate.

To this end, we fit the Green's function results with a function containing algebraic and exponential ($\sim e^{-|i-j|/ \xi}$) decay profiles, and extract the correlation length $\xi$, see Fig.~\ref{figOne}(b) (for details see also \cite{SMPhysRev}).
We consider the difference between the correlation lengths,
$\Delta \xi(T) = \le \xi_{h = 0}(T) - \xi_{h=t}(T) \r$,
in the two regimes at the same temperature $T$ and comparable target fillings $n_t$, for which we know that the charges are confined and deconfined in the ground state, see Fig.~\ref{figOne}(c).
From this we determine the crossover region where thermal fluctuations begin to dominate the exponential decay of the Green's function.
We define the approximate crossover boundary in the region where  $ \xi_{h = 0}(T) - \xi_{h=t}(T) = \frac{1}{2} \xi_{h = 0}(T) $.

We find that the typical crossover region is at ${T/t \approx 0.25}$, which is also influenced by the lattice filling, see Fig.~\ref{figOne}(c).
The so-called target filling $n_t$ is the filling obtained in the ground state at a given chemical potential $\mu$, which is kept constant during the imaginary time evolution.
The actual densities $n(T)$ at finite temperature thus slightly deviate from $n_t$ for each run at $h/t=0$ and $h/t=1$, respectively.
These deviations do not exceed $|n_{h=t}(T) - n_{h=0}(t)| / n(T) < 20\%$ for $T/t < 1$.
We thus plot the data points as a function of $n_t$ \cite{SMPhysRev}.

\textbf{\textit{Friedel oscillations.---}}
Another hallmark of confinement in the $1+1$D \Zt LGT is an abrupt change of the frequency of the Friedel oscillations in the confined phase.
The frequency in the confined phase equals $2 k_F = \pi n$, which is half the frequency found in the deconfined phase of free partons \cite{Borla2020PRL}.
This indicates that the confined mesons are indeed well-defined constituents that remain mobile and form a Luttinger liquid with intricate interactions.

\begin{figure*}[ht]
    \centering
    \epsfig{file=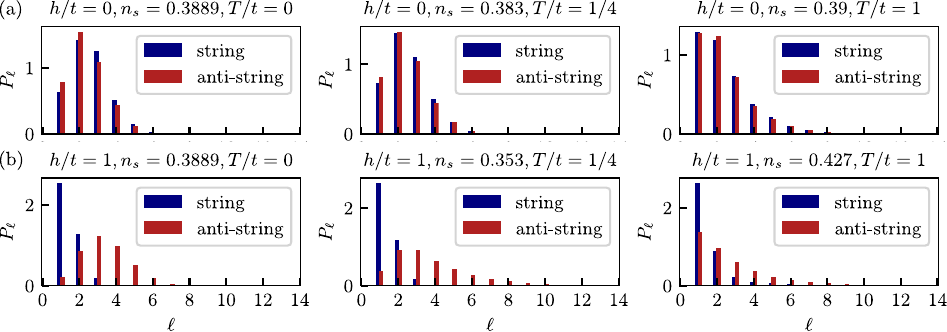}
    \caption{String and anti-string length distributions.
    (a) Distributions are qualitatively similar in the deconfined phase, $h=0$, at a fixed temperature $ T $ and filling obtained from the snapshots $n_s$. The peaks shift to $\ell = 1$ with increasing temperature.
    (b) Different distributions of strings and anti-strings can be observed in the confined phase $h / t = 1$. There is a strong peak at $\ell = 1$ in the string length distribution, and there are no strings with lengths larger then $\ell \geq 3$.
    In contrast, the anti-string length distribution is wide, spreading over $\ell > 8$ at finite temperatures and peaking at around $\ell \approx 3$ in the ground state.
    }
    \label{figThree}
\end{figure*}

In order to analyze the Friedel oscillations at finite temperature, we perform the Fourier transformation of the density profile $\langle \hat{n}_j \rangle$ and extract the frequency of oscillations.
In the deconfined phase $h/t = 0$, we observe broad peaks at $k = 2 \pi n$ which is the expected frequency for the Friedel oscillations of free partons, see Fig.~\ref{figTwo}(a).
The peaks are broad and only become well defined for temperatures $ T / t \leq 1/4$.
With lower temperature the peaks rise and converge to the ground-state results.
Contrarily, we observe peaks at $ k = \pi n$ for low temperatures in the confined phase $h / t = 1$ as expected, see Fig.~\ref{figTwo}(b).
These peaks appear again at around $ T / t \leq 1/4$ and converge to the ground-state results at lower temperature in a similar fashion as in the deconfined case.

There are no deconfined peaks visible in our results for the filling of $n_t = 0.3889$ and $h / t = 1$ at any temperature, which rules out a deconfined parton gas in this regime.
If the later would exist, we would expect a shift in the peak position from $k = \pi n$ to $k = 2 \pi n $ with increasing temperature.
The absence of this shift thus suggests that mesons are pre-formed already well above the crossover temperature, i.e., partons are confined up to high temperatures where thermal fluctuations completely dominate the behavior of the system.

At higher fillings, $n \gtrsim 0.5$, we observe coexistence of peaks at $k = \pi n$ and $2 \pi n $, see Fig.~\ref{figTwo}(c).
However, peaks at both positions rise simultaneously with lower temperature and there is again no exchange of the position of the peaks with temperature.
The peaks observed at higher fillings at $k = 2 \pi n$ can be associated with hole fluctuations, which become significantly more mobile relative to mesons.

\textbf{\textit{String-length distributions.---}}
Our model is within reach of modern cold-atom experiments.
However, extracting the Green's function would be a rather complicated task.
We therefore consider string and anti-string length histograms, that are easily accessible from on-site density-resolved snapshots which can be obtained experimentally.
There, one simply has to extract the number of empty lattice sites between odd-even and even-odd particles respectively, see \cite{SMPhysRev} for more details.
This is a robust, experimentally feasible probe of confinement, since strings are on average shorter than anti-strings in the confined regime; we thus expect different distributions of strings and anti-strings as a clear indicator of confinement.

To demonstrate the effectiveness of such a probe, we sample snapshots from MPS states \cite{Buser2022} using perfect sampling \cite{Ferris2012} implemented withing \textsc{SyTen} \cite{hubig:_syten_toolk, hubig17:_symmet_protec_tensor_network}.
The results presented in Fig.~\ref{figThree} show a clear difference in distributions in the confined and deconfined regimes.
In the deconfined regime there is no difference between the string and anti-string length distributions since partons are free, see Fig.~\ref{figThree}(a).

In the confined phase, the string length distribution is peaked at $\ell = 1$, meaning that most of the mesonic states are tightly confined with few empty lattice sites between the two partons making up a meson, see Fig.~\ref{figThree}(b).
(There is a small fraction of mesons with $\ell \geq 2$, which can be attributed to quantum fluctuations.
The presence of $\ell = 2$ states is in fact necessary for the mesonic states to remain mobile, since the hopping of mesons can be understood as a second-order perturbation process when we consider the limit of $h \gg t$ \cite{Borla2020PRL}.)
In contrast, the anti-string-length distribution is broad, with a long tail.
Furthermore, the anti-string-length distribution has a peak at $\ell > 1$ in the ground state.
This is also influenced by the overall filling of the chain, see \cite{SMPhysRev}.

The combined bimodal distribution of string and anti-string lengths is thus a clear indicator of confinement.
These features are present up to temperatures consistent with our previous calculations of the Green's function and Friedel oscillations.
For higher temperatures $T \geq t$, the distributions become more similar to each other and both peak at $\ell = 1$: this is consistent with a continuous crossover to the deconfined regime at $T = \infty$.
However, at finite temperature $T \gtrsim t$, a slight difference between string and anti-string length histograms remains, supporting our claim of pre-formed mesons up to any finite temperature.

\textbf{\textit{Quench dynamics.---}} Next we consider another experimentally accessible, dynamical probe.
To this end, an initially tightly bound parton pair is introduced into a finite-density thermal gas and  we probe whether it remains confined during the subsequent time evolution.

\begin{figure}[t]
    \centering
    \epsfig{file=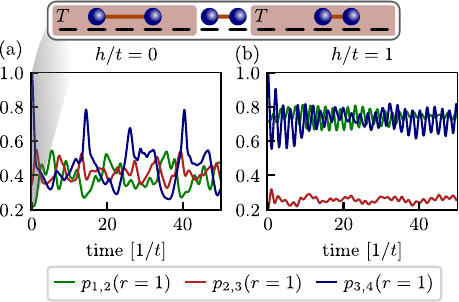}
    \caption{Starting in a thermal ensemble at temperature {$T/t=0.5$} and at half-filling at a given value $h$, with a well-defined particle pair at the middle (see the inset), we quench with $\hat{\mathcal{H}}$ and calculate in ED the probabilities of a pair of particles being one site apart over evolution time. (a) In the case of $h=0$, we find that any two consecutive pairs are equally likely to be a site apart at long evolution times. (b) When $h/t=1$, we find that the middle pair is bound, as are the two other pairs on its either side, which is indicative of confinement.}
    \label{figFour}
\end{figure}

Specifically, we localize a meson on the central two sites of an $L$-site chain; the left (right) remaining $(L-2)/2$ sites are prepared independently in a thermal state of $\H$ in Eq.~\eqref{eqLGTHamiltonian} at a given temperature $T$ and density ${n = (N-2)/(L-2)}$, see inset in Fig.~\ref{figFour}(a).
Then, we calculate the time-evolution of this initial density under the full system Hamiltonian~\eqref{eqLGTHamiltonian}, including all $L$ sites.

We perform numerical exact simulations for $L = 12$, $N=6$ at different values of $h$ and $T$.
Our results indicate no confinement at any temperature when $h=0$, while we again find evidence of confinement at any temperature $T < \infty$ when $h>0$:
We consider dynamics of the probabilities $p_{a,b}(r)$ that the $a^\text{th}$ and $b^\text{th}$ particle, counted from the left, are $r$ sites apart, shown in Fig.~\ref{figFour}(a,b) for $h/t=0$ and $1$, respectively, at $T/t=0.5$.
By construction, the probability of the middle pair to be a site apart is $p_{3,4}(1)=1$ before the quench.
In the wake of the quench, we find a fundamental difference between zero and nonzero $h$. At long times, we find that when $h=0$ any two consecutive particles are equally probable to be a site apart.
On the other hand, when $h/t=1$, we find that it is always more probable that the middle pair is bound, as well as the two pairs to its left and right, indicating confinement.
This qualitative picture holds also at other values of $T$ \cite{SMPhysRev,playlist}, but consistent with a deconfinement crossover, the signal becomes less pronounced for higher temperatures $T \rightarrow \infty$.

\textbf{\textit{Summary and outlook.---}}
In this work, we studied confinement in a $1+1$D \Zt LGT at finite temperature.
We considered a $\mathbb{Z}_2$-invariant Green's function as the direct probe of confinement at finite temperature, where we uncovered a smooth confinement--deconfinement crossover at approximately $T/t \approx 0.25$.
By additionally considering the Friedel oscillations, where the confinement manifests itself in halving of the frequency, we confirmed that the confinement--deconfinment crossover extends up to temperatures where the thermal fluctuations dominate the behavior of the system.
These results were furthermore affirmed by the string and anti-string length distributions that we proposed as an experimentally feasible, robust measure of confinement.
Finally, we complemented our results with dynamical probes, also experimentally readily accessible in current state-of-the-art quantum simulators \cite{Kuno2017}.
There, we showed that, again, confinement persists up to high temperatures, albeit signatures of confinement become less pronounced as the system approaches the deconfined infinite-temperature state.

Our results pave the way towards understanding confinement crossover at finite temperature in a simple $1+1$D \Zt LGT with dynamical matter, which can be probed with current quantum simulators.
We show that the partons remain confined at low temperature, with a smooth crossover at finite temperature to an incoherent regime dominated by thermal fluctuations.
At any finite temperature $T < \infty$, signatures of confinement remain.
This result challenges the conventional reasoning in one-dimension, where one would naively expect a deconfined regime at any finite temperature as the system is deconfined for $T \rightarrow \infty$.
We expect that our results can be extended to higher gauge groups and models with more complicated interactions.
Our work paves the way for explorations of confinement in state-of-the-art analog, or digital, quantum simulators, which naturally include thermal fluctuations.
Such setups can also naturally explore mixed-dimensional settings of coupled $1$D chains, where even richer confinement-deconfinement physics can be expected \cite{Grusdt2020}.

\begin{acknowledgments}
We thank Annabelle Bohrdt, Zohreh Davoudi, Lukas Homeier, Mattia Moroder, Henning Schlömer, Alexander Schuckert, and Christopher White for fruitful discussions.
This research was funded by the Deutsche Forschungsgemeinschaft (DFG, German Research Foundation) under Germany's Excellence Strategy -- EXC-2111 -- 390814868 
and via Research Unit FOR 2414 under project number 277974659, 
and received funding from the European Research Council (ERC) under the European Union’s Horizon 2020 research and innovation programm (Grant Agreement no 948141) — ERC Starting Grant SimUcQuam.
J.C.H. and F.G.~acknowledge funding within the QuantERA II Programme that has received funding from the European Union’s Horizon 2020 research and innovation programme under Grand Agreement No 101017733, support by the QuantERA grant DYNAMITE and by the Deutsche Forschungsgemeinschaft (DFG, German Research Foundation) under project number 499183856.

\end{acknowledgments}

%


\clearpage
\pagebreak
\newpage
\onecolumngrid
\widetext
\begin{center}
\textbf{\large Supplemental Material: Confinement in 1+1D \texorpdfstring{$\mathbb{Z}_2$}{Z2} Lattice Gauge Theories at Finite Temperature}
\end{center}

\setcounter{equation}{0}
\setcounter{figure}{0}
\setcounter{table}{0}
\setcounter{page}{1}
\makeatletter
\renewcommand{\theequation}{S\arabic{equation}}
\renewcommand{\thefigure}{S\arabic{figure}}

\section{Numerical simulations of the ground state}
\begin{figure}[ht]
\centering
\epsfig{file=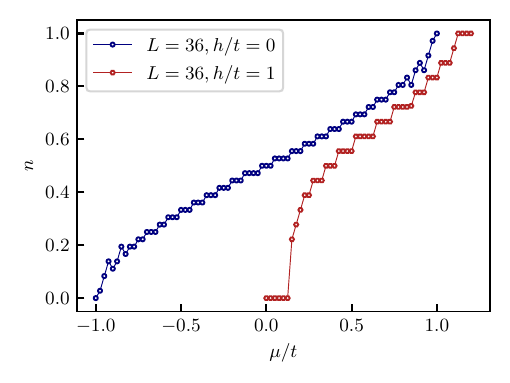}
\caption{Lattice filling $n$ as a function of chemical potential $\mu$ in the absence of field $h = 0$ (blue) and at finite field $h / t = 1$ (red) in the ground state. Total chain length is $L = 36$.
}
\label{FigSupOne}
\end{figure}

Ground-state calculations are performed using finite-system DMRG \cite{Schollwoeck2011, White1992} through the DMRG toolkit \textsc{SyTen} \cite{hubig:_syten_toolk, hubig17:_symmet_protec_tensor_network}.
Furthermore, we use the Gauss-law constraint and map the original \Zt lattice gauge theory (LGT) Hamiltonian to the pure spin-$1/2$ Hamiltonian \cite{Borla2020PRL, Kebric2021, KebricNJP2023}
\begin{equation}
    \H_s = t \sum_{j = 1}^{L-1} \left( 4 \hat{S}_{j-1}^{x}\hat{S}_{j+1}^{x} - 1 \right) \hat{S}_{j}^{z}
    - 2 h \sum_{j=0}^{L} \hat{S}_{j}^{x}
    + 4 \mu \sum_{j=0}^{L-1} \hat{S}_{j}^{x}\hat{S}_{j+1}^{x},
    \label{eqLGTConstrainedSpinModel}
\end{equation}
where $\mu$ is the chemical potential that we add in order to control the filling $n$, see Fig.~\ref{FigSupOne}.
The devil's staircase structure comes from the fact that we perform our calculations on a chain of finite length.
Hence, the width of the observed plateaus of constant filling are proportional to the charge gap, which is system size-dependent.
This gap completely disappears in the thermodynamic limit $L \rightarrow \infty$ in the considered parameter regime presented in Fig.~\ref{FigSupOne}.

The mapping of the $1+1$D \Zt LGT to the spin model \eqref{eqLGTConstrainedSpinModel} is exact and comes from the Gauss-law constraint, where we only consider the physical sector defined as $\hat{G}_j \ket{\psi} = + 1 \ket{\psi}, \forall j$ \cite{Prosko2017}.
This constraint explicitly relates spin configurations to the position of hard-core bosons as $\hat{n}_j = \frac{1}{2} \le 1 - 4 \hat{S}_{j}^{x}\hat{S}_{j+1}^{x} \r$ \cite{Kebric2021}.
Note that the total number of spin sites is equal to $L+1$, where $L$ is the number of matter lattice sites, since the chain always begins and ends with a link.
The Green's function presented in the main text can also be rewritten in terms of spin operators and reads
$ \mathcal{G}(i-j) = \left \langle 
    \le \prod_{i \leq \ell \leq j} 2 \hat{S}_{ \ell}^{z} \r
    \frac{1}{2} \le 1 - 4 \hat{S}_{i-1}^{x}\hat{S}_{i}^{x} \r
    \frac{1}{2} \le 1 + 4 \hat{S}_{j}^{x}\hat{S}_{j+1}^{x} \r \right \rangle $.
In addition, we denote the total number of hard-core bosons in the chain as $N$ and the filling is thus defined as $n = N / L$.
We typical simulate chains up to $L = 36$.
We could easily increase the chain length up to $L \gtrsim 100$ for the ground-state calculations; however, such lengths would become increasingly difficult to compute at finite temperature.
For easier comparison with the finite-temperature calculations we thus limit our calculations to lower system sizes.

\section{Finite-temperature simulations}
Finite temperature calculations are performed using the purification scheme where we enlarge our Hilbert space by adding an auxiliary lattice site to every physical lattice site \cite{Feiguin2005, Feiguin2013, Nocera2016}.
Here a thermal state is represented with a pure state of the extended system as \cite{Feiguin2005, Nocera2016}
\begin{equation}
    \ket{\psi (\beta)} = e^{-\beta \H / 2} \ket{\psi(\beta = 0)},
    \label{eqThermalStateDef}
\end{equation}
where $\beta = 1 / T$ is the inverse temperature $T$ and $\ket{\psi(\beta = 0)}$ is a maximally entangled state between physical and auxiliary lattice sites.
Thermodynamic averages of physical observables are computed as \cite{Nocera2016, Feiguin2005}
\begin{equation}
    \left \langle \mathcal{\hat{O}} \right \rangle =
    \frac{\bra{\psi(\beta)} \mathcal{\hat{O}} \ket{\psi(\beta)}}{ \bra{\psi(\beta)} \ket{\psi(\beta)} }.
    \label{eqPurifiedObsevable}
\end{equation}

\begin{figure}[ht]
\centering
\epsfig{file=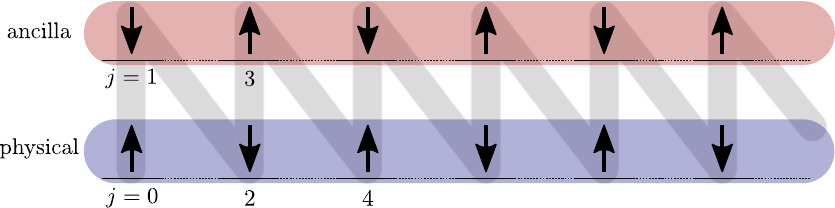, width=0.65\textwidth}
\caption{Physical and auxiliary (ancilla) lattice sites of the enlarged system where $j$ denotes the index of the MPS lattice site.
}
\label{FigSupTwo}
\end{figure}

By attaching an auxiliary lattice site to every physical lattice site, we double our spin chain length $L + 1 \rightarrow 2(L + 1)$, which we implement with matrix product states (MPS), see also Fig.~\ref{FigSupTwo}.
Furthermore, we consider physical sites to reside on even lattice sites and for the auxiliary lattice sites to reside on odd lattice sites as proposed in \cite{Feiguin2005}.
In order to implement the maximally entangled state $\ket{\psi(\beta = 0)}$ between physical and auxiliary sites, we first use DMRG to calculate the ground state of the \textit{entangler} Hamiltonian \cite{Nocera2016}
\begin{equation}
    \H_e = - \sum_{j = 0}^{L} \le \hat{S}^{+}_{2j} \hat{S}^{-}_{2j+1} + \hc \r .
    \label{eqLGTEntangler}
\end{equation}
To be more precise, the resulting state is that where physical lattice sites are maximally entangled to their corresponding auxiliary lattice sites.
Auxiliary lattice sites can thus be understood as providing a thermal bath \cite{Feiguin2005}.

Finally, in order to obtain finite-temperature states, we perform imaginary time evolution, Eq.~\eqref{eqThermalStateDef}, of our MPS \cite{Paeckel2019}, with the LGT Hamiltonian \eqref{eqLGTConstrainedSpinModel} that acts only on the physical lattice sites and is rewritten as 
\begin{equation}
    \H_s^{t} = t \sum_{j = 1}^{L-1} \left( 4 \hat{S}_{2j-2}^{x}\hat{S}_{2j+2}^{x} - 1 \right) \hat{S}_{2j}^{z}
    - 2 h \sum_{j=0}^{L} \hat{S}_{2j}^{x}
    + 2 \mu \sum_{j=0}^{L-1} \hat{S}_{2j}^{x}\hat{S}_{2j+2}^{x}.
    \label{eqLGTConstrainedSpinModelFiniteT}
\end{equation}
We use the Krylov algorithm for the first few time steps and the TDVP algorithm for the remaining time steps \cite{Paeckel2019}, which are both implemented in \textsc{SyTen} \cite{hubig:_syten_toolk, hubig17:_symmet_protec_tensor_network}.
The initial maximally entangled state $\ket{\psi(\beta = 0)}$ has a very low bond dimension of only $\chi = 2$ on every other MPS lattice site.
This is the reason why we us the Krylov algorithm for the first $10$ time steps of $\Delta \beta_K t / 2 = 0.01$ in order to increase the bond dimension in a controlled way.
For the successive time evolution, we use the two-site TDVP algorithm with time step of $\Delta \beta_T t / 2 = 0.05$ all the way up to the final inverse temperature $\beta t / 2 = 10$.
We typically limit the truncation error below $\sim 10^{-9}$ per time-step.

\begin{figure}[t]
\centering
\epsfig{file=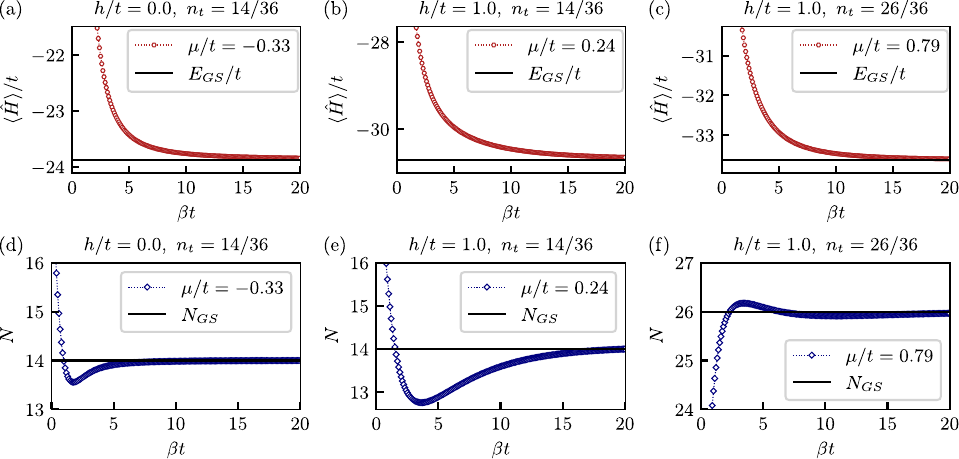}
\caption{(a) Expectation value of the Hamiltonian \eqref{eqLGTConstrainedSpinModel} at a chemical potential $\mu / t = 0.33275$ which yields the ground-state filling of $n = 14/36$ in the deconfined regime $h / t =0 $.
(b) Expectation value of the Hamiltonian in the confined regime $h / t = 1$ at a chemical potential $\mu / t = 0.242$, which also gives the filling of $ n = 14 /36 $ in the ground state.
The convergence to the ground state is slightly slower in the confined regime.
(c) Finally, we also present the expectation value of the Hamiltonian for a chemical potential $\mu /t = 0.6875$, which yields a slightly higher target filling of $n_t = 24 / 36 $ in the confined regime.
We observe a monotonic decrease of the energy with increasing inverse temperature $\beta = 1 / T$ in all cases.
Horizontal black lines denoted with $E_{GS}/t$ are the values of the ground-state energies at the corresponding chemical potential.
Expectation value of the total particle number $N$ as a function of $\beta = 1 / T$ at a chemical potential with the ground-state filling of $n = 14/36$ in the deconfined regime $h / t =0 $ (d) and in the confined regime $h / t = 1$ at a chemical potential which yields the same target filling of $ n_t = 14 /36 $ (e), as well as the particle number at a chemical potential with a slightly higher target filling of $n_t = 26 / 36 $ in the confined regime (f).
The expectation values of the particle number slightly overshoot the target fillings for low values of $\beta$ in all cases and then quickly converge to the ground-state results with increasing $\beta$.
Horizontal black lines denoted with $N_{GS}$ are the corresponding target values in the ground state.
}
\label{FigSupThree}
\end{figure}

In order to benchmark the results, we compute the expectation value of the Hamiltonian and the total number of particles $N$ as a function of inverse temperature $\beta$, see Fig.~\ref{FigSupThree}.
Both quantities converge towards the target values obtained from the ground-state calculations with increasing $\beta$.
The expectation value of the Hamiltonian monotonically decreases towards the ground-state results since the total energy of the system decreases as it cools down, see Fig.~\ref{FigSupThree}(a)--(c).
The expectation value of the total particle number has a more interesting behavior.
It typically slightly overshoots the ground-state target at $\beta t \approx 2$ and then converges to the integer target value with increasing $\beta$ as expected, see Fig.~\ref{FigSupThree}(d)--(f).

In all cases we use a constant value of the chemical potential $\mu$ while performing the imaginary time evolution.
The choice of $\mu$ was made by considering the devil's staircase structure of the ground-state calculations presented in Fig.~\ref{FigSupOne}.
Hence our target filling becomes more precise for higher values of inverse temperature.
However, the biggest error of the average filling that we get for $\beta t > 1$ is at most around $\Delta N \sim \pm 2$.
This results in a density error of around $\Delta n \sim 20 \%$ 
for lowest filling $n = 0.2778$ presented in the main text.
This error decreases dramatically for higher fillings.
We could obtain more precise results by varying the chemical potential value for specific temperature.
This would be a rather tedious task where we would have to map out different chemical potentials and their resulting fillings at different temperatures.
However, since the errors are nevertheless relatively low and most importantly, well controlled, we consider such an approach not necessary.
Our results capture all the qualitative features and we are not interested in extracting very precise numerical values in great detail.

\begin{figure}[ht]
\centering
\epsfig{file=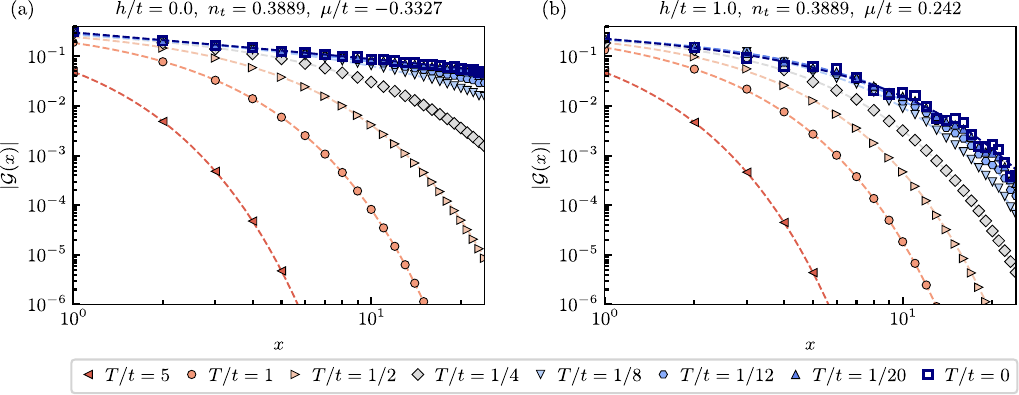}
\caption{Log-log plots of the Green's function at a target filling of $n_t = 0.3889$ for different temperatures and electric field strengths.
(a)~Green's function in the absence of the electric field term $h = 0$, where ground-state results clearly exhibit a linear decay in the log-log plot, which corresponds to the algebraic decay attributed to the deconfined phase.
Finite-temperature results in this regime start to deviate from the ground-state calculations with higher temperature $T $.
(b)~For finite electric field strength $h / t = 1$, the Green's function decays exponentially for any finite temperature.
The decay is slower for low temperatures where the results converge to the ground-state results.
The most obvious change in the exponential decay in both cases presented here can be observed for temperatures $T / t \geq 4$.
The actual densities at finite temperature follow the curves presented in Fig.~\ref{FigSupThree}(a)--(c).
}
\label{FigSupFour}
\end{figure}
\section{Green's function fits}
As mentioned in the main text, we want to compare the exponential decay of the Green's function at finite temperature in the deconfined regime to the confined regime.
To this end, we fit the absolute value of the Green's function results obtained from our numerical calculations with a simple function which contains exponenital and algebraic decay,
\begin{equation}
    f_G = A x^{- \alpha}e^{ - x / \xi}.
    \label{eqFitFctGreens}
\end{equation}
Here we defined the correlation length $\xi$ of the exponential decay and a power-law decay exponent $\alpha$.
Due to the exponential nature of the numerical results, we fit the logarithm of our data.
Hence, we rewrite the fitting function defined above in Eq.~\eqref{eqFitFctGreens} as 
\begin{equation}
    \tilde{f}_G = \tilde{A} - \alpha \log(x) - \frac{x }{\xi}.
    \label{eqFitFctLogGreens}
\end{equation}
Example fits in the log-log scale are presented in Fig.~\ref{FigSupFour}.
The finite-temperature results clearly converge to the ground-state calculations as the temperature decreases.
This is seen in the deconfined case presented in Fig.~\ref{FigSupFour}(a).
Ground-state calculations exhibit a power-law decay which is reflected in a linear curve in the log-log plot.
Slight deviations from the power-law decay can already be seen for the lowest-temperature data sets.
The biggest difference can then be seen for the data set at temperature $T / t = 1/4$, where the exponential decay is already pronounced.
This is also consistent with the onset of Friedel oscillations discussed in the main text (see also the section on Friedel oscillations in the supplementary material).
Similar convergence to the ground-state results with lower temperature $T$ can also be observed in the case when the confining electric field is non-zero, $h / t = 1$, see Fig.~\ref{FigSupFour}(b).
The onset of deviations from the ground-state results can also be observed at around $T / t \sim 1/4$, which is again consistent with the Friedel-oscillation results in the main text.

We compare the extracted correlation lengths at $h/t = 0$ to the results at $h / t = 1$ at the same temperature $T$ and at approximately same filling $n$ by computing the difference between the correlation lengths in both regimes 
$\Delta \xi(T) = \le \xi_{h = 0}(T) - \xi_{h=t}(T) \r $, which is presented in the main text.
The actual fillings $n$ at finite temperature in both regimes are slightly off from the target filling $n_t$ in the ground state and follow the curves similar to those presented in Fig.~\ref{FigSupThree}(d)--(f).
To be more precise, every set of vertical data points positioned at a constant $n_t$ in Fig.~1(c) of the main text, comes from comparing results for $h/t = 0$ and $h/t = 1$ at constant chemical potentials $\mu$ in the both regimes, which give the corresponding $n_t$ in the ground state.
This results in errors of the horizontal position of the data points which we discuss in the previous section of the supplementary material, as well as in errors of the actual correlation lengths. 
However, the errors are relatively small and the qualitative behavior is still captured.
We therefore take the target filling value obtained in the ground state when presenting our results in the main text as extrapolating between the precise fillings $n(h /t = 0, T)$ and  $n(h /t = 1, T)$ would unnecessarily complicate the general picture.

\section{Friedel oscillations}
\begin{figure}[ht]
\centering
\epsfig{file=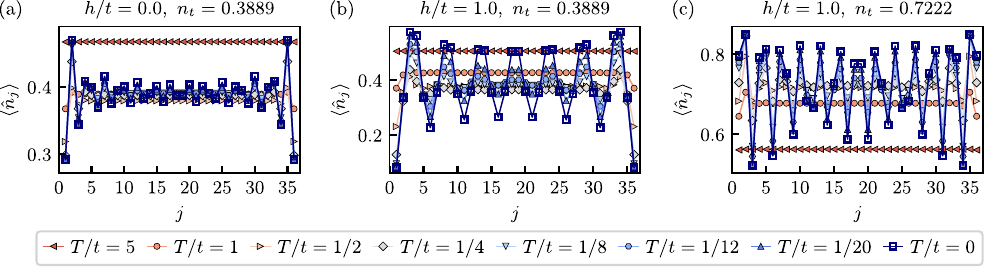}
\caption{Density profile for different target fillings in the confined and deconfined regimes.
(a) Friedel oscillations for filling $n_t = 14 / 36 \approx 0.3889$ in the deconfined regime $h = 0$.
(b) The frequency of the Friedel oscillations in the confined regime when $h / t = 1$ is half the frequency in the deconfined regime at the same filling. This can be easily observed by comparing the number of peaks in (a) and (b).
(c) Friedel oscillations at higher filling $n_t = 26/36 = 0.7222$ in the confined regime.
In both cases, the finite-temperature Friedel oscillations converge to the ground-state results with decreasing temperature $T$.
The actual densities at finite temperature follow the curves presented in Fig.~\ref{FigSupThree}(a)--(c).}
\label{FigSupFive}
\end{figure}

\subsection{Real space density profile at finite temperature}
Here we briefly comment on the density profiles in the chain for different fillings in the confined and deconfined regimes which we present in Fig.~\ref{FigSupFive}.
In the main text, we study the behavior of Friedel oscillations at finite temperature by considering the Fourier transformation to extract their frequencies.
By doing so, we determine up to which temperature signatures of confinement persist.
Here we show that the same behavior is already apparent by considering the Friedel oscillations directly.
At high temperatures, we observe a featureless flat density profile and
Friedel oscillations appear only when the temperature is lowered.
This is the same in the confined and deconfined regimes regardless of the filling, see Fig.~\ref{FigSupFive}.
First signatures of oscillations appear close to the edges of the system at around $T / t \approx 1/2$.
Oscillations become stronger and visible also in the bulk with decreasing temperature $ T / t < 1/4$ when the oscillations start to resemble the ground-state results.

By comparing the Friedel oscillations at the same target filling $n_t = 0.3889$ in the deconfined and confined regimes in Fig.~\ref{FigSupFive}(a) and (b), one can observe that the frequency of oscillations in the confined regime is indeed only half the frequency in the deconfined regime.
We also show oscillations at a higher filling $n_t = 0.7222$ where we observe double peaks in the Fourier transform.
Although the oscillatory behavior becomes slightly less clear due to the high frequency, we do see that the leading frequency is the one corresponding to the confined phase.

\subsection{Fourier transformation of the Friedel oscillations}
The Fourier transform of the Friedel oscillations which we present in the main text is defined as
\begin{equation}
    n_k = \frac{1}{L} \sum_{j = 0}^{L-1} e^{-ikj}
    \left \langle  \hat{n}_j \right \rangle.
    \label{eqFourierOfFriedel}
\end{equation}
We discretize our $k$ modes as $\Delta k = \frac{2 \pi }{L}$.

\section{String and anti-string length distributions from snapshots}
\begin{figure}[ht]
\centering
\epsfig{file=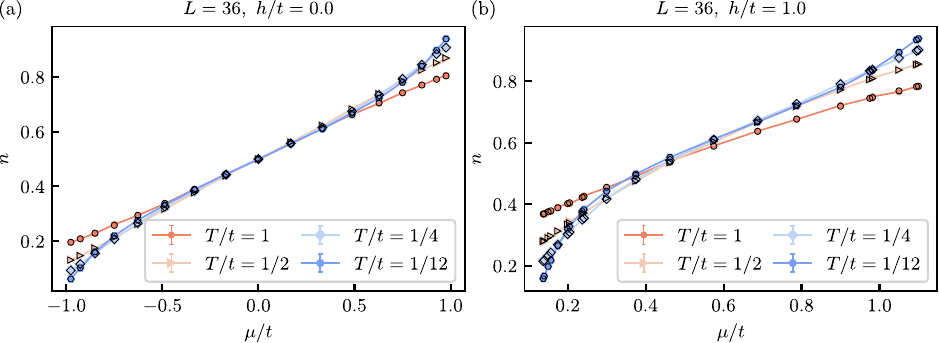}
\caption{Lattice filling for different temperatures as a function of chemical potential obtained from the snapshots (filled markers with error bars) and filling results obtained directly from the MPS calculations (black empty markers).
(a) Density profiles are symmetric in $\mu = 0$ as expected in the deconfined regime $h/t = 0$.
(b) Slight asymmetry is observed in $\mu$ in the confined regime $h/t = 1$.
The same marker shapes represent the same temperature of the snapshot and MPS calculation in all cases.
The ground-state results match almost perfectly across different chemical potentials and temperatures in the confined and deconfined regimes.
The error bars which we define as $\sigma / \sqrt{N_s}$, where $\sigma $ is the standard deviation and $N_s$ is the number of snapshots, are smaller than marker size.
}
\label{FigSupSix}
\end{figure}
As already mentioned in the main text, we sample snapshots from MPS \cite{Buser2022} using the so-called perfect sampling \cite{Ferris2012}.
The algorithm for sampling snapshots is implemented in the \textsc{SyTen} toolkit \cite{hubig:_syten_toolk, hubig17:_symmet_protec_tensor_network}.
We sample snapshots in the $x$-basis of our spin-$1/2$ chain.
In each snapshot we thus obtain the configuration of the \Zt electric fields $\hat{\tau}^{x}$ on every lattice site.
We then extract the length of every string and anti-string in the snapshot.
This is done by considering the distances between odd-even and even-odd particles, respectively.
To locate the particles in our chain, we once again use the Gauss law, where we consider the physical sector that yields the simple connection between the spin configuration on the links and particle number on the sites $\hat{n}_j = \frac{1}{2} \le 1 - 4 \hat{S}_{j}^{x}\hat{S}_{j+1}^{x} \r$.
Hence, we simply search for the domain walls in the spin configuration in order to extract the positions of particles.
We typically sample $2000$ snapshots from every MPS to produce the histograms in the main text.

We also check the average density $n_s$ of partons obtained from snapshot sampling and compare them with the results obtained directly from the MPS, see Fig.~\ref{FigSupSix}.
The snapshot results in the ground state exactly match the results obtained from the MPS directly.
Finite-temperature results match the particle number on average, since we performed grand canonical calculations at finite temperature, i.e., we also have snapshots where the particle number is slightly lower or higher than the target filling.
As a result we also have snapshot contributions with an odd number of particles, which do not break the Gauss law.
A small statistical error is acquired also due to the finite number of taken snapshots.

\section{Dynamical calculations}

\begin{figure}[ht]
\centering
\epsfig{file=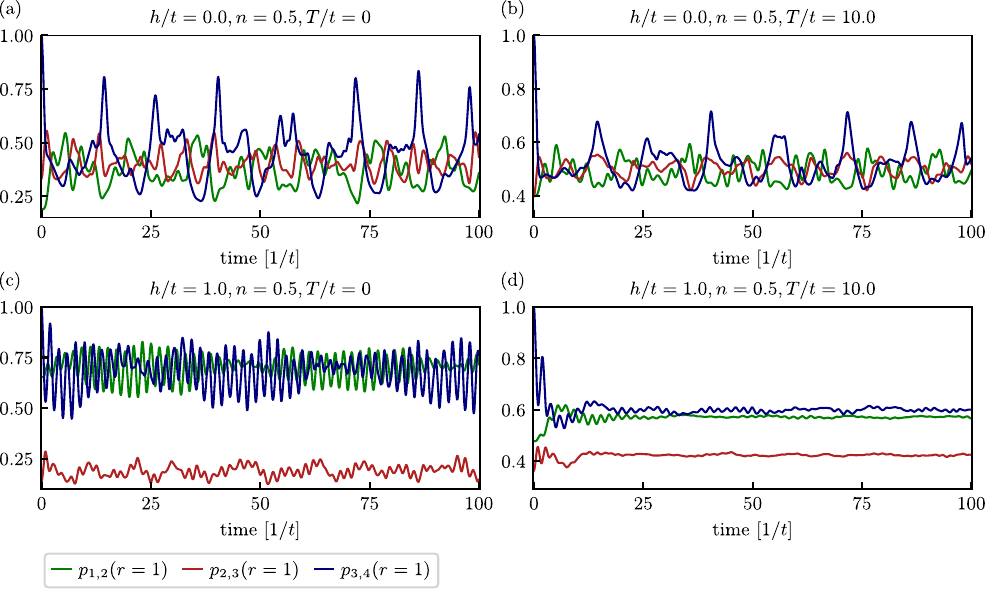}
\caption{Starting in a thermal ensemble at temperature $T$ and at half-filling at a given strength $h$ of the electric field, with a well-defined particle pair at the middle sites and an equal number of particles to its left and right, we quench with $\hat{\mathcal{H}}$ and calculate in ED the probabilities of a pair of particles being one site apart over evolution time. In the case of $h=0$, we find that there is no confinement regardless of temperature, as shown in (a) for $T=0$ and in (b) for $T/t=10$. When $h=t$, we find that at long times there will always be signs of confinement regardless of temperature, as shown in (c) for $T=0$ and in (d) for $T/t=10$.
}
\label{FigSupSeven}
\end{figure}

With experimental feasibility in mind, let us consider the initial state

\begin{align}\label{eq:InitialEnsemble}
    \hat{\rho}(0)=\hat{\rho}_\text{L}(0)\otimes\ketbra{n_{\frac{L}{2}}=1}\otimes\ketbra{\tau^x_{\frac{L}{2},\frac{L}{2}+1}=-1}\otimes\ketbra{n_{\frac{L}{2}+1}=1}\otimes\hat{\rho}_\text{R}(0),
\end{align}
where $\hat{\rho}_{\text{L}(\text{R})}(0)=e^{-\hat{\mathcal{H}}_{\text{L}(\text{R})}/T}$ and $\hat{\mathcal{H}}_{\text{L}(\text{R})}$ describe the original Hamiltonian $\hat{\mathcal{H}}$, Eq.~(1) in the main text, on the left (right) $L/2-1$ sites and links in the filling sector $L/4-1$ and gauge sector $g_j=+1,\,\forall j$. The initial state~\eqref{eq:InitialEnsemble} is experimentally easy to prepare. It involves pinning a meson pair in the center of the chain, while letting the left and right parts of the system thermalize independently at temperature $T$, e.g., by coupling to an approximate thermal bath at temperature $T$.

We then quench this initial state with $\hat{\mathcal{H}}$ to obtain the time-evolved density operator

\begin{align}
    \hat{\rho}(t)=e^{-i\hat{\mathcal{H}}t}\hat{\rho}(0)e^{i\hat{\mathcal{H}}t}.
\end{align}
In our ED calculations, we have used open boundary conditions with $L=12$ matter sites.

We then calculate the dynamics of the probabilities $p_{a,b}(r)$ that the $a^\text{th}$ particle from the left is $r$ sites apart from the $b^\text{th}$ particle, again counting from the left edge of the system.
Our ED results indicate a fundamental difference between $h=0$ and $h\neq0$.
In all cases, the probability of the two middle particles to be a site apart starts at unity by construction of the initial state~\eqref{eq:InitialEnsemble}. However, at long times we find that it is roughly equally probable for any two consecutive particles to be a site apart when $h=0$, regardless of the temperature, as shown in Fig.~\ref{FigSupSeven}(a,b) for $T/t=0$ and $10$, respectively.

On the other hand, when $h=t$, we find that at late times it is always more probable that the initially bound particles will remain close to one another, as shown in Fig.~\ref{FigSupSeven}(c,d) for $T/t=0$ and $10$, respectively. We have also tried even higher temperatures (not shown), and this picture always holds, although the signal becomes less pronounced when $T\to\infty$. We also provide \href{https://www.youtube.com/playlist?list=PLoUsb3eaKix5yAeQWXmCgnU88Ivn9BzsR}{videos} of the time evolution of the parton-separation probabilities.

\end{document}